# Side-channel attack on labeling CAPTCHAs

Carlos Javier Hernandez-Castro[1], Arturo Ribagorda[1], Yago Saez[2]

[1]Security Group, Department of Computer Science, Carlos III University, 28911 Leganes, Madrid, Spain
[2]EVANNAI Group (Artificial Neural Networks and Evolutionary Computation), Carlos III University, Computer Science Department, Madrid, Spain
{chernand, arturo}@inf.uc3m.es, yago.saez@uc3m.es

**Abstract.** We propose a new scheme of attack on the Microsoft's ASIRRA CAPTCHA which represents a significant shortcut to the intended attacking path, as it is not based in any advance in the state of the art on the field of image recognition. After studying the ASIRRA Public Corpus, we conclude that the security margin as stated by their authors seems to be quite optimistic. Then, we analyze which of the studied parameters for the image files seems to disclose the most valuable information for helping in correct classification, arriving at a surprising discovery. This represents a completely new approach to breaking CAPTCHAs that can be applied to many of the currently proposed image-labeling algorithms, and to prove this point we show how to use the very same approach against the HumanAuth CAPTCHA. Lastly, we investigate some measures that could be used to secure the ASIRRA and HumanAuth schemes, but conclude no easy solutions are at hand[1].

## 1 Introduction

The last decade has seen increasing interest in abusing services provided in the Internet, mainly for economical reasons. There has been misuse of services like e-mail account creation for spam sending and phishing, abuse of sites where anonymous posting is encouraged (Wikipedia, blogs comments, news sites, etc.) for adding links for commercial promotion, harassment or vandalism. There has also been abuse of remote voting mechanisms[14]. Automatic (script-guide) site wandering has also been described as a way to facilitate resource consumption and thus remote denial-of-service attacks. Anonymous abuse of on-line games, inclusive for commercial promotion, is not new. Other denial-of-service attacks on public posting sites (like employment listings or CV reception e-mails addresses) is also possible. Thus, there are lots of sounding economical reasons to abuse services provided through the Internet.

The main trend to prevent this automatic abuse has been to develop the ability to tell humans and computers apart - remotely and through an untrust-worthy channel. Many tests -generically called CAPTCHAs[2] or HIPs[3]- have been developed with that aim. Those tests rely on capacities inherent to the human mind but supposedly difficult to mimic for computers, that is, problems that have been traditionally hard to solve in computers (as problems that still remain wide open for Artificial Intelligence researchers).

---
[1] This article is part of a research work still in progress

Moni Naor seems to have been the first to propose theoretical methods of telling apart computers from humans remotely to prevent the abuse of web services in [5]. In 1997, primitive CAPTCHAs were developed by Andrei Broder, Martin Abadi, Krishna Bharat, and Mark Lillibridge to prevent bots from adding URLs to their search engine [6]. The term CAPTCHA (for Completely Automated Turing Test To Tell Computers and Humans Apart) was coined in 2000 by Luis von Ahn, Manuel Blum, Nicholas Hopper and John Langford of Carnegie Mellon University [3]. At the time, they developed the first CAPTCHA to be used by Yahoo. Those earlier designs were mostly text-based: the computer chose a random sequence of letters and rendered them in an image after applying different kinds of distortions. The human challenger, supposedly far better than a computer in character recognition, was to identify the characters. But even after graphic distortion and degradation, some approaches have been able to "read" them and thus solve the test automatically around 92% of the time [4], specially so when it is possible to divide the graphic into its constituent letters. Some approaches have focussed on making this division harder, typically to the expense of making it also harder to the human challenger.

There are CAPTCHAs that rely on the same foundations as text CAPTCHA but seem to be slightly stronger, but also more difficult for the common user, so can only be used as special-purpose CAPTCHAs for certain types of human clients. Among them there is MAPTCHA, the Mathematical Captcha, that shows a math formula (typically a limit) and asks the user to write down its numerical value solution. This CAPTCHA, apart from being probably not difficult to attack by advanced OCR, has other shortcomings in its current form (i.e.: the answer requested has to be numeric).

### 1.1 Image CAPTCHAs

General vision seems to be a harder problem than character recognition, so more designs have focused on using pictures instead - even though most of those CAPTCHAs do not really rely on a "general vision problem" but in a downsized version of categorizing images.

Chew and Tygar[7] were the first to use a set of labeled images to produce CATPCHAs challenges. For that purpose, they used the labeled associated with images in Google Image Search. This technique is not well suited for CAPTCHAs, as Google relates a picture to its description and its surroundings, so the word 'bicycle' could refer to a bicycle or to a music band name. Ahn and Dabbish[8] proposed a new way to label images by embedding the task as a game, called the "ESP game". However, it has a fixed number of object classes (70) and the image database seems not large enough. The site HotCaptcha.com proposed a way to use a large-scale human labeled database provided by the HotOrNot.com website, a site that invites users to post photos of themselves and rate others' in a numerical scale of "hotness". It is true that beauty is influenced by culture and subjective, but still there is a large consensus on some beautiful attributes. This proposal is no longer active, and as of Jan-2009 the site HotCaptcha.com is down. Oli Warner came with the idea of using photos of kittens to tell computers and humans apart [9]. KittenAuth features nine pictures of cute little animals, only three of which are feline. The problem with this system is that its database of pictures is small enough ($< 100$) to manually classify them, and this limitation seems troublesome even if we apply new methods involving image distortion. ASIRRA [1] uses a similar approach but using a giant database of more that 3 million photos from Petfinder.com, a website devoted to finding homes for homeless pets, with a daily addition of around 10,000 more pics. Asirra displays 12 images from the database (mostly composed of dogs or cats images) and asks the user to select the cats in it. As it is using Petfinder.com web service, it provides a link for adopting each pet, promoting the aim of Petfinder.com of finding new pet owners.

---

[2] Completely Automated Public Turing test to tell Computers and Humans Apart
[3] Human Interactive Proof

### 1.2 Motivation

We do not fully understand yet if all the problems chosen and in use in the design of current CAPTCHAs are really hard enough, and if their design and implementation could be error-prone and make them easier for automatic solving. It is then interesting to try to break current CAPTCHAs and find pitfalls in their design to make the state-of-the-art advance and get to a point when well known and tested assumptions give base for more secure CAPTCHAs.

### 1.3 Organization

The rest of the paper is organized as follows: In the next section, we introduce the ASIRRA CAPTCHA in greater detail. After this, in Section 3 we describe the ENT tool, which will be helpful in the attack against ASIRRA described in Section 4. Then, in Section 5 we examine other works that have investigated the ASIRRA scheme and explain the differences with our work. In Section 6 we explain how our new scheme can be used against almost any other type of image CAPTCHAs and show a demonstration involving HumanAuth, and finally, in Section 7 we extract some conclusions and propose possible improvements together with future research lines.

## 2 The ASIRRA CAPTCHA

The ASIRRA CAPTCHA[1] is based on the task of identifying images under two categories, cats and dogs (ASIRRA stands for "Animal Species Image Recognition for Restricting Access"). A challenge consists on a set of 12 photographs, all the same pixel size, each of which represents a cat or a dog. To solve the CAPTCHA the user has to highlight all the photographs depicting cats and none depicting dogs. According to [1], ASIRRA "can be solved by humans 99.6% of the time in under 30 seconds", implying a significant advantage in usability compared with character-based CAPTCHAs.

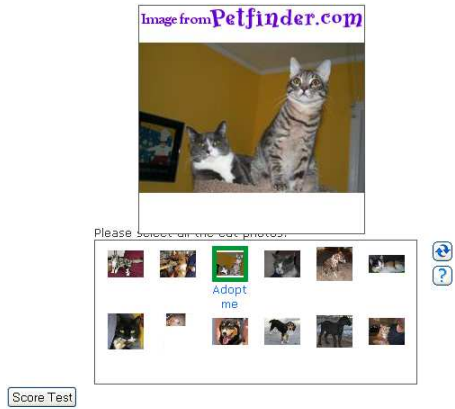

**Fig. 1.** ASIRRA CAPTCHA in action

ASIRRA has some additional features due to its partnertship with PetFinder.com. It displays pets for adoption that are close to the IP geolocation of the client receiving the CAPTCHA, and below each image it displays an "adopt me" link that redirects the client to a PetFinder.com page with more information about the pet. ASIRRA developers have nonetheless put great care in these two features not affecting the security of their proposal. For example, when an user selects an "adopt me" link, the CAPTCHA is invalidated before being redirected to PetFinder.com. They have also put some extra effort in the usability, introducing schemes to prevent the issuing of difficult to classify images (images with more than one pet, or pets of two different races, etc.).

For testing its security, ASIRRA authors have developed a classifier based on histograms of color features described in [1] that is only 56.9% accurate. The authors state [1] that "based on a survey of machine vision literature and vision experts at Microsoft Research, we believe classification accuracy of better than 60% will be difficult without a significant advance in the state of the art".

## 3 Description of the ENT tool

The ENT[11] tool is a program comprising a suite of different tests that search for information density and randomness in a byte sequence. It does so applying various tests to the sequence of bytes and reporting the numerical results of those tests. As these tests are useful for evaluating information quantity and randomness quality, this results are of interest for evaluating pseudo-random number generators and studying the output of compression algorithms. In the novel way presented in this paper, they can also be used for file/image classification.

The tests included in the ENT program are:

- Entropy: information density of the contents of the file, expressed as the mean number of bits necessary to represent a character of 8 bits (byte). For example, if the last 4 bits of each byte are always 0110 and the former 4 bits vary 'randomly', then the mean entropy of the file would be 4 (4 bits per byte).
- Compression: this test tells us the size shrink (in percentage) we could obtain if the file was compressed using a lossless compression algorithm of the Lempel-Ziv type (one pass).
- Chi-square test: this tests computes the expected p-value for a distribution with 256 degrees of liberty (dividing the file in 8 bit chunks of data). This p-value represents how frequently a uniform distribution would exceed the computed value.
- Arithmetic mean: this tests computes the mean value of all the bytes of the input file.
- Monte-Carlo value for Pi: this tests uses a Monte-Carlo (probabilistic) algorithm to compute the value of Pi, using the input file as the source of randomness for such algorithm.
- Serial correlation: this test measures how a byte of the file can be approximated by its preceding byte.

For an idea of the results one can expect of the ENT test, we show the output obtained when applied to the following input files:

- An ASCII English version of Don Quijote de la Mancha, by Miguel de Cervantes, from the Gutemberg Project [4]
- A BMP non-compressed image (posted on the Internet) [5]
- A WAV non-compressed sound (posted on the Internet) [6]
- A JPEG compressed image from Flickr.com [7]
- The Chase, an MP3 compressed music file from the Internet Audio Open Source Archive [8]

| $test$ | $ASCII$ | $BMP$ | $WAV$ | $JPEG$ | $MP3$ |
|---|---|---|---|---|---|
| size | 2347772 | 1683594 | 116904 | 4914423 | 2916331 |
| entropy | 4.49 | 7.24 | 6.16 | 7.91 | 7.85 |
| compression | 43 % | 9 % | 22 % | 1 % | 1 % |
| chi square | 37,346,041.42 | 2,796,525.96 | 484,762.37 | 669,239.83 | 1,643,472.32 |
| chi square p-value | > 0.01 | > 0.01 | > 0.01 | > 0.01 | > 0.01 |
| mean | 88.92 | 73.21 | 125.74 | 137.88 | 119.89 |
| Monte-Carlo Pi | 4.00 | 3.45 | 3.97 | 2.84 | 3.20 |
| serial correlation | 0.016057 | 0.537042 | 0.928775 | 0.004862 | 0.168447 |

The entropy and compression tests give information about data density. The mean and Monte Carlo Pi tells us about value distribution, the serial correlation about data interrelation (also related to data redundancy), and the chi square test is the most sensible to non random data distribution.

---

[4] http://www.gutenberg.org/files/996/996.txt
[5] http://www.lossip.com/wp-content/uploads/marc-anthony-y-ricardo-arjona.bmp
[6] http://amazingsounds.iespana.es/oceanwaves.wav
[7] http://farm4.static.flickr.com/3126/3153559748_$b0ee7fd24b\_o.jpg$
[8] http://www.archive.org/download/Green-LiveAtTheAQ/TheChase.mp3

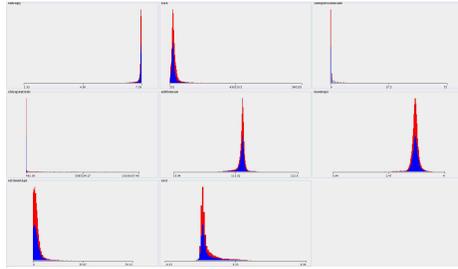

**Fig. 2.** Distribution of classes (different colours) regarding ENT test results

## 4  Attack against ASIRRA

We have downloaded the ASIRRA Public Corpus, that the developers of ASIRRA have generously created to help researchers study the security of their proposal. It is composed of around 25.000 images, classified into two directories named Cat and Dog, with one half of the images in each class. We then analyzed all of the jpeg files contained herein with the help of the ENT tool, producing a formatted output (in ARFF format, so to be used with Weka [10]) with its results, after discarding two corrupt files. This were later processed by a classifier which was able of distinguishing cats and dogs pictures with a nearly 60% accuracy, without using any kind of image recognition technique.

An especially simple, easy-to-understand classifier is shown in Figure 1 below:

```
=== Run information ===
Scheme:       weka.classifiers.meta.AttributeSelectedClassifier
              -E "weka.attributeSelection.CfsSubsetEval "
              -S "weka.attributeSelection.BestFirst -D 1
              -N 5" -W weka.classifiers.trees.J48 -- -C 0.25 -M 2
Relation:     catsdogs
Instances:    24998
Attributes:   9
              entropy
              size
              compressionrate
              chisqstatistic
              arithmean
              montepi
              errmontepi
              corr
              class
Test mode:    10-fold cross-validation

=== Classifier model (full training set) ===

AttributeSelectedClassifier:

Classifier Model
J48 pruned tree
------------------

size <= 32137: Cat (15998.0/7013.0)
size > 32137: Dog (9000.0/3514.0)

=== Stratified cross-validation ===
=== Summary ===

Correctly Classified Instances       14448               57.7966 %
Incorrectly Classified Instances     10550               42.2034 %
Total Number of Instances            24998

=== Confusion Matrix ===

    a    b   <-- classified as
 8969 3530 |   a = Cat
```

```
7020 5479 |    b = Dog
```

In the figure above we can see how the simplest decision tree[13] based on only the **size** of a jpeg file is able of distinguishing between cats and dogs with an accuracy well over 57% over the 24998 images, so significantly better than random.

A more complex classifier could perform slightly better, reaching a correctly classified ratio of 58.0326 with a LogitBoost[12] technique with 10 boost iterations and a DecisionStump as the underlying algorithm (see below). These results are averages taken after a 10-fold cross-validation process.

```
=== Run information ===
Scheme:       weka.classifiers.meta.LogitBoost
              -P 100 -F 0 -R 1 -L -1.7976931348623157E308 -H 1.0 -S 1 -I 10
              -W weka.classifiers.trees.DecisionStump
Relation:     catsdogs
Instances:    24998
Attributes:   9
              entropy
              size
              compressionrate
              chisqstatistic
              arithmean
              montepi
              errmontepi
              corr
              class
Test mode:    10-fold cross-validation

=== Classifier model (full training set) ===

LogitBoost: Base classifiers and their weights

=== Stratified cross-validation ===
=== Summary ===

Correctly Classified Instances       14507              58.0326 %
Incorrectly Classified Instances     10491              41.9674 %
Total Number of Instances            24998

=== Confusion Matrix ===

    a    b   <-- classified as
 8288 4211 |    a = Cat
 6280 6219 |    b = Dog
```

In conclusion, what we have here is that even a very simple and efficient classifier based on a completely side characteristic of the image files (in this case, its size) is able of telling apart cats from dogs with a much better than random accuracy.

This is clearly an undesirable property of the ASIRRA scheme, and, for that matter, of any CAPTCHA scheme, as if correctly designed they are expected to be broken only if some major advance in artificial intelligence is achieved.

### 4.1 ASIRRA Attribute Selection

A natural question that might arise is which of the output values computed by the ENT tool over the image files are most informative for a side-channel classification. As can be easily deduced from the simplest classifier shown in the last section, it seems that the value of the **size** parameter is the most relevant, but what about the others? For discovering the most meaningful parameters for cat and dog image classification, we have performed an attribute selection process with the Weka [10] classification tool, using a chi-square based evaluator with a Ranker search method, as shown below:

```
=== Run information ===

Evaluator:    weka.attributeSelection.ChiSquaredAttributeEval
Search:       weka.attributeSelection.Ranker -T -1.7976931348623157E308 -N -1
Relation:     catsdogs
Instances:    24998
Attributes:   9
[.....]
Evaluation mode:    10-fold cross-validation

=== Attribute selection 10 fold cross-validation (stratified), seed: 1 ===

average merit      average rank  attribute
729.115 +-23.245   1    +- 0      2 size
223.755 +-19.423   2    +- 0      4 chisqstatistic
171.656 +-17.037   3    +- 0      5 arithmean
144.709 +-10.347   4    +- 0      1 entropy
 81.333 +- 6.923   5.1  +- 0.3    8 corr
 73.121 +- 4.512   6    +- 0.45   6 montepi
 66.109 +-10.686   6.9  +- 0.3    7 errmontepi
 16.297 +- 8.621   8    +- 0      3 compressionrate
```

As expected, the first attribute in terms of information disclosure merit is size, with around 729.115, or 3.25 times more relevance than the chi-square statistic which is in second place, performing slightly better than the arithmetic mean that, with around 171.656 is a little more informative than file entropy. All the rest seem to be not too helpful for telling dogs and cats apart, especially the compression rate.

One can think that this disclosure of classification information by the size parameter should be very easy to repair. The ASIRRA developers can easily redesign the image filter applied to the images taken from Petfinder.com, and process them to have exactly the same size when forming an ASIRRA challenge, so that the size of all proposed images is identical and, consequently, it reveals no useful information whatsoever. Unfortunately, this seems not to be the case. If we remove the size parameter completely, the classification is slightly worse but still much better than random, as shown below:

```
=== Run information ===

Scheme:       weka.classifiers.meta.AttributeSelectedClassifier
              -E "weka.attributeSelection.CfsSubsetEval "
              -S "weka.attributeSelection.BestFirst -D 1
              -N 5" -W weka.classifiers.trees.J48 -- -C 0.25 -M 2
Relation:     catsdogs-weka.filters.unsupervised.attribute.Remove-R2
Instances:    24998
Attributes:   8
[.....]
Test mode:    10-fold cross-validation

=== Classifier model (full training set) ===

AttributeSelectedClassifier:

=== Attribute Selection on all input data ===

Search Method:
    Best first.
    Start set: no attributes
    Search direction: forward
    Stale search after 5 node expansions
    Total number of subsets evaluated: 29
    Merit of best subset found:    0.009

Attribute Subset Evaluator (supervised, Class (nominal): 8 class):
    CFS Subset Evaluator
    Including locally predictive attributes

Selected attributes: 1,3,4 : 3
                     entropy
                     chisqstatistic
                     arithmean

Classifier Model
J48 pruned tree
------------------
```

```
chisqstatistic <= 1330.01
|   chisqstatistic <= 1173.43: Cat (4864.0/1981.0)
|   chisqstatistic > 1173.43
|   |   entropy <= 7.971873: Cat (2084.0/905.0)
|   |   entropy > 7.971873: Dog (627.0/286.0)
chisqstatistic > 1330.01
|   entropy <= 7.965789
|   |   arithmean <= 122.5622
|   |   |   chisqstatistic <= 5825.38
|   |   |   |   entropy <= 7.933784
|   |   |   |   |   entropy <= 7.718804
|   |   |   |   |   |   arithmean <= 104.4152: Cat (11.0/2.0)
|   |   |   |   |   |   arithmean > 104.4152: Dog (43.0/11.0)
|   |   |   |   |   entropy > 7.718804: Cat (1358.0/482.0)
|   |   |   |   entropy > 7.933784
|   |   |   |   |   chisqstatistic <= 1848.76: Cat (684.0/259.0)
|   |   |   |   |   chisqstatistic > 1848.76: Dog (670.0/324.0)
|   |   |   chisqstatistic > 5825.38
|   |   |   |   entropy <= 7.923559: Cat (2709.0/1260.0)
|   |   |   |   entropy > 7.923559: Dog (64.0/18.0)
|   |   arithmean > 122.5622
|   |   |   chisqstatistic <= 2020: Cat (3969.0/1962.0)
|   |   |   chisqstatistic > 2020: Dog (3749.0/1626.0)
|   entropy > 7.965789: Dog (4166.0/1406.0)

=== Stratified cross-validation ===
=== Summary ===

Correctly Classified Instances       14218               56.8766 %
Incorrectly Classified Instances     10780               43.1234 %
Total Number of Instances            24998
```

Here, the same classifier that was able of obtaining a 57.7966% using all the attributes is still able of achieving a 56.8766% without the size parameter, although it struggles to do that, and is forced to construct a much larger and complex tree than in the previous case.

After discarding the size value, the most informative parameter is the chi-square value. If we remove both the size and the chi-square value, the rest of the set of ENT output values are still able of producing a better-than-random classification (with around a 54.1763% accuracy over the 24998 images of the Corpus).

Although modifying the chi-square value of the image files to remove it as we have done below is a much harder task than that corresponding to the size parameter, it additionally seems not to be the solution to avoid this kind of side-channel attacks against the ASIRRA CAPTCHA.

```
=== Run information ===
Scheme:       weka.classifiers.meta.LogitBoost
              -P 100 -F 0 -R 1 -L -1.7976931348623157E308
              -H 1.0 -S 1 -I 10 -W weka.classifiers.trees.DecisionStump
Relation:     catsdogs-weka.filters.unsupervised.attribute.Remove-R2,4
Instances:    24998
Attributes:   7
[.....]
Test mode:    10-fold cross-validation

=== Classifier model (full training set) ===

LogitBoost: Base classifiers and their weights:

=== Stratified cross-validation ===
=== Summary ===

Correctly Classified Instances       13543               54.1763 %
Incorrectly Classified Instances     11455               45.8237 %
Total Number of Instances            24998
```

| classifier | non-allowed parameters | parameters used | accuracy |
|---|---|---|---|
| LogitBoost/DecisionStump | - | all | 58.0326 % |
| J48 pruned tree | - | size | 57.7966 % |
| J48 pruned tree | size | entropy, chi-square, mean | 56.8766 % |
| LogitBoost/DecisionStump | size, chi-square | rest | 54.1763 % |

## 5  Other Works

In Philippe Golle's recent work [2], he presents the strongest attack against ASIRRA as of Jan-2009 (still unpublished). This attack is based in image processing, as it divides the photographs into NxN cells of color and texture (grayscale) information, and use that to feed two support-vector machine (SVM) classifiers that, when used together, are capable of classifying with around a 83% accuracy, thus allowing them to solve the 12-photos challenge with a 10.3% probability.

For that purpose they collected 13,000 images from the ASIRRA implementation (publicly available at the ASIRRA website) avoiding duplicates (they detected 6). Then they classified them into Cat (49.3%), Dog (49.7%) and Other (1.0% - pictures with no recognizable animal, or either both cat and dog). Then the authors experimented with different color and texture features extracted from the images. Using those features, they trained support-vector machines (SVM) classifiers using a 5-fold cross validation (dividing a subset of images into 5 partitions, 4 used for training and 1 for validation) using subsets of various sizes (reporting average results through 5 experiments with subsets of 5,000 and 10,000 images).

For color processing, the authors subdivide the color space in regions. They subdivide the HSV (hue, saturation, value) model of color in $C_h$, $C_s$ and $C_v$ bands of equal width. Then, for each NxN cell of the image, they create a vector with a 1 in the region if there is at least one pixel with color in that region. This approach is boolean-based (recalls on the presence or not of the color) and not frequency-based (like the color-histogram approach tried by the ASIRRA authors [1]). They find that color-presence features are more accurate for classifying cats and dogs than color-histogram features. To explain this counter-intuitive result, they hypothesize that color-presence features are scale independent (and not color-histogram), and also that the distribution of color-presence features is much more regular [2] than the distribution of real-valued color histograms. With this scheme, the authors reach an 77.1% accurate classification rate based on a training set of 8,000 images.

For texture processing, the authors experiment with an statistical approach based in intensity measures in different regions of the image and with a structural approach, in which a texture is defined as a set of texture tiles in repeated patterns, finding the later to be more useful for the creation of classifiers. In this approach, they define a set of textures (of 5x5 pixels) and create a feature vector of each image with distances from the image to each one of this textures (normalized in the range [0,1]). Using this method, the authors reach an 80.4% accurate classification rate based on a training set of 8,000 images. Combining both methods, they reach an 82.7% accurate classification rate.

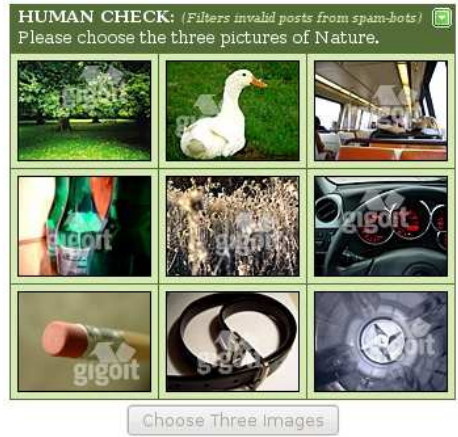

**Fig. 3.** HumanAuth CAPTCHA capture with the GigoIt logo as an example watermark

This approach is quite interesting and produces an impressive outcome, but it is rather conventional as it relies in actual imaging processing techniques, extracting color information and texture patterns, calculating distances between textures, and using that information in the classification phase, thus effectively advancing the state-of-the-art in image recognition (at least in the cat/dog instance).

Our approach is completely different: it does not use any technique related to image processing, but categorizes files by other means and is also likely to be much more effective in terms of speed, and more general as can be applied to any image-based CAPTCHA.

Also, as our approach is completely different from the classical approach presented in [2], both can be combined to obtain an improved classification ratio.

## 6 Other CAPTCHAs affected

In this section we describe the results of this side-channel analysis technique against other interesting CAPTCHA proposal called HumanAuth (http://sourceforge.net/projects/human

The HumanAuth CAPTCHA is based on the ability of humans to distinguish between images with natural and non-natural contents. The source code of the HumanAuth application comes with a image repository consisting of 45 nature images and 68 non-nature ones in jpeg format. The idea is quite interesting, and the CAPTCHA is specially easy for humans and purportedly very difficult for non-humans.

## 6.1 HumanAuth image repository

Following an identical approach to that used in the ASIRRA case, we passed the ENT tool to all the images in the HumanAuth source code, producing an ARFF file for Weka processing. The best classifier (in this case RandomForest) was able to show an accuracy rate of 77.8761% (see below), which is significantly better than the $\frac{68}{68+45} = 60.177\%$ that a trivial classifier (that always predicts the larger class) will do.

```
=== Run information ===

Scheme:       weka.classifiers.trees.RandomForest -I 10 -K 0 -S 1
Relation:     humanauth
Instances:    113
Attributes:   9
[.....]
Test mode:    10-fold cross-validation
Random forest of 10 trees, each constructed while considering 4 random features.
=== Summary ===
Correctly Classified Instances          88              77.8761 %
Incorrectly Classified Instances        25              22.1239 %

=== Confusion Matrix ===
  a  b   <-- classified as
 34 11 |  a = nature
 14 54 |  b = nonnature
```

In the case of the HumanAuth CAPTCHA, the information disclosure of the output ENT values is distributed much more evenly, except for the values related with the MonteCarlo estimation of $\pi$ which seem to disclose absolutely no useful data for the classification. This implies that devising measures to difficult side-channel attacks against this scheme is even harder than in the already very difficult case of the ASIRRA CAPTCHA.

```
=== Run information ===

Evaluator:    weka.attributeSelection.ChiSquaredAttributeEval
Search:       weka.attributeSelection.Ranker -T -1.7976931348623157E308 -N -1
Relation:     humanauth
Instances:    113
Attributes:   9
[.....]
Evaluation mode:    10-fold cross-validation

=== Attribute selection 10 fold cross-validation (stratified), seed: 1 ===

average merit      average rank  attribute
38.987 +- 5.975    1.9 +- 1.58   8 corr
34.58  +- 3.872    2.6 +- 1.2    3 compressionrate
32.936 +- 6.025    2.7 +- 0.78   2 size
29.844 +- 4.878    3.9 +- 0.83   1 entropy
29.885 +- 4.57     4.6 +- 1.85   4 chisqstatistic
27.228 +- 1.393    5.3 +- 0.46   5 arithmean
 0     +- 0        7.2 +- 0.4    7 errmontepi
 0     +- 0        7.8 +- 0.4    6 montepi
```

Even after completely removing the three most significant attributes for the HumanAuth classifier, that is correlation, compression rate and size, an SMO algorithm is able of achieving a 75.2212% accuracy.

```
=== Run information ===

Scheme:       weka.classifiers.functions.SMO -C 1.0 -L 0.0010
              -P 1.0E-12 -N 0 -V -1 -W 1
              -K "weka.classifiers.functions.supportVector.PolyKernel -C 250007 -E 1.0"
Relation:     humanauth-weka.filters.unsupervised.
              attribute.Remove-R8-weka.filters.unsupervised.
              attribute.Remove-R3-weka.filters.unsupervised.
              attribute.Remove-R2
Instances:    113
Attributes:   6
              entropy
              chisqstatistic
```

```
           arithmean
           montepi
           errmontepi
           class
Test mode:    10-fold cross-validation

=== Classifier model (full training set) ===

SMO

Kernel used:
  Linear Kernel: K(x,y) = <x,y>

Classifier for classes: nature, nonnature

BinarySMO

Machine linear: showing attribute weights, not support vectors.

         0.9365 * (normalized) entropy
 +      -2.1654 * (normalized) chisqstatistic
 +      -0.1669 * (normalized) arithmean
 +       0.0142 * (normalized) montepi
 +       0.0527 * (normalized) errmontepi
 +       0.2322

Number of kernel evaluations: 2089 (76.363% cached)

=== Stratified cross-validation ===
=== Summary ===

Correctly Classified Instances          85               75.2212 %
Incorrectly Classified Instances        28               24.7788 %
Total Number of Instances              113

=== Confusion Matrix ===

  a  b   <-- classified as
 17 28 |  a = nature
  0 68 |  b = nonnature
```

| classifier | non-allowed parameters | parameters used | accuracy |
|---|---|---|---|
| RandomForest | - | all | 77.8761 % |
| SMO | correlation, compression rate, size | rest | 75.2212 % |

It seems, therefore, extremely difficult to protect these set of images against the proposed side-channel analysis, which seriously casts a doubt over the security of the derived HumanAuth CAPTCHA.

## 6.2 HumanAuth CAPTCHA watermarking

To prevent easy image library indexing, the authors of the HumanAuth CAPTCHA decided to merge a constant PNG image with the random JPG image taken from the library. To do that, it locates the PNG in a random position into the JPG canvas and merges both using a level of transparency, so the PNG appears as a watermark, and does not distort the JPG image as much as to make it difficult for the human eye to recognise the image. The HumanAuth source includes one PNG image (a logo) that can be used for testing.

We have created a set of 20,000 images, 10,000 of each class (nature and non-nature) and have extracted statistical information from them with the ENT tool, building an ARFF file for Weka processing. We have obtained the following results.

```
=== Run information ===

Scheme:       weka.classifiers.meta.LogitBoost -P 100 -F 0 -R 1 -L -1.7976931348623157E308 -H 1.0 -S 1 -I 10 -W weka.classifiers.trees.DecisionStump
Relation:     humanauth
Instances:    20000
Attributes:   9
[.....]
Test mode:    10-fold cross-validation
```

```
=== Classifier model (full training set) ===
LogitBoost: Base classifiers and their weights:
[.....]
=== Summary ===
Correctly Classified Instances       14517               72.585 %
Incorrectly Classified Instances      5483               27.415 %
Kappa statistic                          0.4517
Mean absolute error                      0.3631
Root mean squared error                  0.4229
Relative absolute error                 72.6115 %
Root relative squared error             84.5869 %
Total Number of Instances            20000
[.....]
=== Confusion Matrix ===

    a    b   <-- classified as
 8093 1907 |    a = nature
 3576 6424 |    b = nonnature
```

This result can be somewhat expected: with the little randomness introduced by merging the logo in a random position, the precision of the classifier drops to 72.585 %.

```
=== Run information ===

Scheme:       weka.classifiers.trees.J48 -C 0.25 -M 2
Relation:     humanauth
Instances:    20000
Attributes:   9
[.....]
Test mode:    10-fold cross-validation
=== Classifier model (full training set) ===
J48 pruned tree
------------------
[.....]
Number of Leaves  :  481
Size of the tree :   961
[.....]
=== Summary ===
Correctly Classified Instances       18187               90.935 %
Incorrectly Classified Instances      1813                9.065 %
Kappa statistic                          0.8187
Mean absolute error                      0.1113
Root mean squared error                  0.2776
Relative absolute error                 22.2538 %
Root relative squared error             55.5252 %
Total Number of Instances            20000
[.....]
```

Using a classification tree of 481 leaves, J48 is able to reach almost a 91 % accuracy. This is due to the "repetition" of images, as we have created an image set of 20,000 images which are statistically close (enough) to the 45 original ones. This suggests that the initial small set of images, when used with the scheme proposed by the HumanAuth authors of merging with a watermark, may not be of use against this type of attack, even though might be enough to prevent hash-function (like MD5) indexing - which is the intention of the designer. One can argue that we can choose a different watermark that alters more the original image, but that would be also at the expense of human visual recognition. It can be argued that other possible approach could be randomly using a set of different watermarks, but that would be at the expense of creating an appropriate set, so we are just moving/distributing the original problem (images can be characterized because of a not enough uniform distribution).

As can be expected when using as a seed such a small set of images, the classification accuracy raises as enough samples are given to build a good enough decision tree, reaching its top logarithmically. The figures given here correspond to another test set we have created:

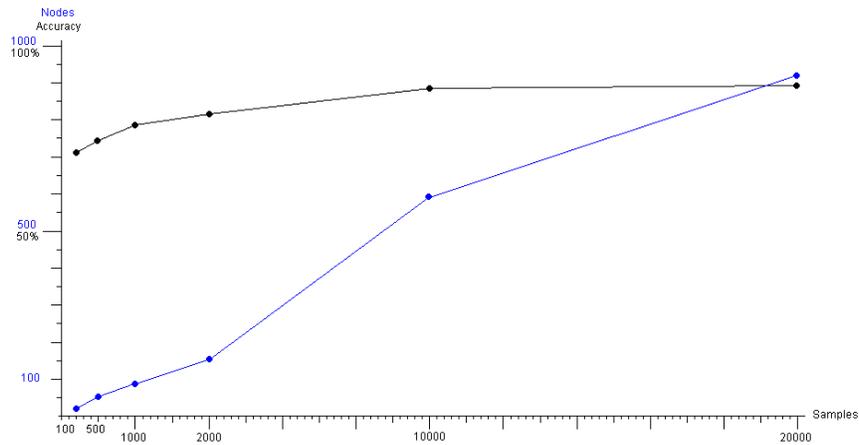

**Fig. 4.** Increase in accuracy and number of nodes with images available

| Images | Nodes | Accuracy |
|--------|-------|----------|
| 200    | 21    | 71.5 %   |
| 500    | 61    | 74.4 %   |
| 1000   | 85    | 78.6 %   |
| 2000   | 159   | 82.1 %   |
| 10000  | 597   | 88.8 %   |
| 20000  | 929   | 89.7 %   |

Other classification schemes show even better results than J48:

```
=== Run information ===

Scheme:       weka.classifiers.trees.RandomForest -I 10 -K 0 -S 1
Relation:     humanauth
Instances:    20000
Attributes:   9
[.....]
Test mode:    10-fold cross-validation
=== Classifier model (full training set) ===
Random forest of 10 trees, each constructed while considering 4 random features.
Out of bag error: 0.1207
=== Summary ===
Correctly Classified Instances       18538               92.69  %
Incorrectly Classified Instances      1462                7.31  %
Kappa statistic                          0.8538
Mean absolute error                      0.1133
Root mean squared error                  0.2327
Relative absolute error                 22.658 %
Root relative squared error             46.5399 %
Total Number of Instances            20000
[.....]
```

Theese results are still good after the most significant attributes are removed from the classification scheme.

| classifier | non-allowed parameters | parameters used | accuracy |
|------------|------------------------|-----------------|----------|
| LogitBoost/DecisionStump | - | all | 72.585 % |
| J48 | - | all | 90.935 % |
| RandomForest | - | all | 92.69 % |
| RandomForest | correlation, compression rate, size | rest | 82.805 % |

# 7 Concluding Remarks

We address in the following, Section 7.1, the generality of the presented attack, and later some conclusions and ideas for future works.

## 7.1 Attack Generality

Our approach can be used as a very general analysis tool to realistically estimate the security parameters of any CAPTCHA proposal, and we believe it will be advisable to use it in the future before any similar systems are launched to have adequate, well-reasoned, and founded security parameters and realistic estimations. One of their main advantages is that it does not depend on the underlying format (image, sound, video, etc.) or problem, and that it could be useful for avoiding pitfalls such as the existence of some trivial and irrelevant parameter values (i.e. size) leaking too much class-relevant information.

## 7.2 Conclusions and Future Work

The ASIRRA CAPTCHA is very interesting, but as our work and this only previous work [2] have shown, their security estimates are too optimistic. Additionally, the credit scheme proposed by the authors should not be recommended (a conclusion also presented in [2]) because it increases a lot the attacker probability of getting a ticket.

The lessons learned in this analysis are useful to improve other attacks based on more common approaches -like image processing- or, alternatively, can be used to improve the security of these CAPTCHA schemes, and this could be an interesting future work. For example, filtering images taken from Petfinder to force then to have a much similar average size and less standard deviation could harden the task of the attacker without affecting the overall good properties of the ASIRRA proposal.